\documentclass[conference]{IEEEtran}
\IEEEoverridecommandlockouts

\usepackage{cite}
\usepackage{amsmath,amssymb,amsfonts}
\usepackage{algorithmic}
\usepackage{graphicx}
\usepackage{textcomp}
\usepackage{xcolor}
\def\BibTeX{{\rm B\kern-.05em{\sc i\kern-.025em b}\kern-.08em
    T\kern-.1667em\lower.7ex\hbox{E}\kern-.125emX}}
\begin{document}

\newcommand\grl{Geophys.~Res.~Lett.}
\newcommand\apjl{ApJ}
\newcommand\aap{A\&A}
\newcommand\apj{ApJ}
\newcommand\apjs{ApJS}
\newcommand\ao{Appl.~Opt.}
\newcommand\nat{Nature}
\newcommand\pasp{PASP}
\newcommand\ssr{Space Science Reviews}
\newcommand\solphys{Sol.~Phys.}
\newcommand\mnras{Monthly Notices of the Royal Astronomical Society }
\newcommand{\add}{\bf \color{blue}}
\newcommand{\delete}{\st}

\title{Out-of-Sample Validation of MagNet 

}

\author{

\IEEEauthorblockN{Aryiadna Yesmanchyk}
\IEEEauthorblockA{\textit{Department of Physics} \\
\textit{New Jersey Institute of Technology}\\
Newark, New Jersey \\
ay338@njit.edu}
\and
\IEEEauthorblockN{Yan Xu}
\IEEEauthorblockA{\textit{Institute for Space Weather Sciences} \\
\textit{New Jersey Institute of Technology}\\
Newark, New Jersey \\
yan.xu@njit.edu
}
\and
\IEEEauthorblockN{Jason T. L. Wang}
\IEEEauthorblockA{\textit{Institute for Space Weather Sciences} \\
\textit{New Jersey Institute of Technology}\\
Newark, New Jersey \\
wangj@njit.edu
}
\and
\centering
\IEEEauthorblockN{Haodi Jiang}
\IEEEauthorblockA{\textit{Department of Computer Science} \\
\textit{Sam Houston State University}\\
Huntsville, Texas \\
haodi.jiang@shsu.edu}

\and
\IEEEauthorblockN{Chunhui Xu}
\IEEEauthorblockA{\textit{Institute for Space Weather Sciences} \\
\textit{New Jersey Institute of Technology}\\
Newark, New Jersey \\
cx4@njit.edu}

\and
\IEEEauthorblockN{Haimin Wang}
\IEEEauthorblockA{\textit{Institute for Space Weather Sciences} \\
\textit{New Jersey Institute of Technology}\\
Newark, New Jersey \\
haimin.wang@njit.edu}
}

\maketitle

\begin{abstract}

Machine learning is starting to be used in almost every industry and academic research, and solar physics is no exception. A newly developed machine learning model named MagNet \cite{Jiang2023} helps us to tackle some of the most serious challenges in data mining by generating transverse fields of solar active regions. Being trained on line-of-sight magnetograms from Michelson Doppler Imager at Solar and Heliospheric Observatory (SOHO/MDI), H$\alpha$ maps from Big Bear Solar Observatory and Kanzelh{\"o}he Solar Observatory and vector magnetograms from Helioseismic and Magnetic Imager at Solar Dynamic Observatory (SDO/HMI), this model provides vector magnetograms in active regions for SOHO/MDI data covering the strong solar cycle 23. In this study, we performed out-of-sample validation of the MagNet model with data from Imaging Vector Magnetograph (IVM) at Mees Solar Observatory, which was not included in the training process. Our results show good correlation between the AI generated data and the observed vector magnetograms and therefore strengthen the confidence of implementing MagNet to the entire SOHO/MDI archive and future scientific analysis of the AI generated data. 

 
\end{abstract}

\section{Introduction}

It has been well known that the topology and evolution of magnetic fields are crucial in providing energy storage and release of solar eruptive events. Magnetic field parameters, such as free energy, are derived from the photospheric vector magnetograms and used to study solar flares and CMEs, and eventually help in predicting solar eruptions. The most
recent solar cycle 24 was relatively weak with few large flares, though it is the only solar cycle in
which consistent time-sequence vector magnetograms have been available through the Helioseismic and Magnetic Imager \cite{HMI} on board the Solar Dynamics Observatory
\cite{SDO} since its launch in 2010. On the other hand, the Michelson Doppler
Imager \cite{MDI} on board the Solar and Heliospheric Observatory \cite{SOHO} from 1996 to 2011 observed stronger events in Solar Cycle 23. However, MDI only collected line-of-sight (LOS) magnetograms, which are not suitable for comprehensive analysis of solar active regions (ARs). This created a need for a machine learning model that could be used to generate these missing but crucial data. We are specifically interested in reconstructing B$x$ and B$y$ components, as B$z$ can be directly taken from the LOS magnetogram and does not require reconstruction.

\begin{figure*}[pht]
\centering
\includegraphics[scale=0.68]{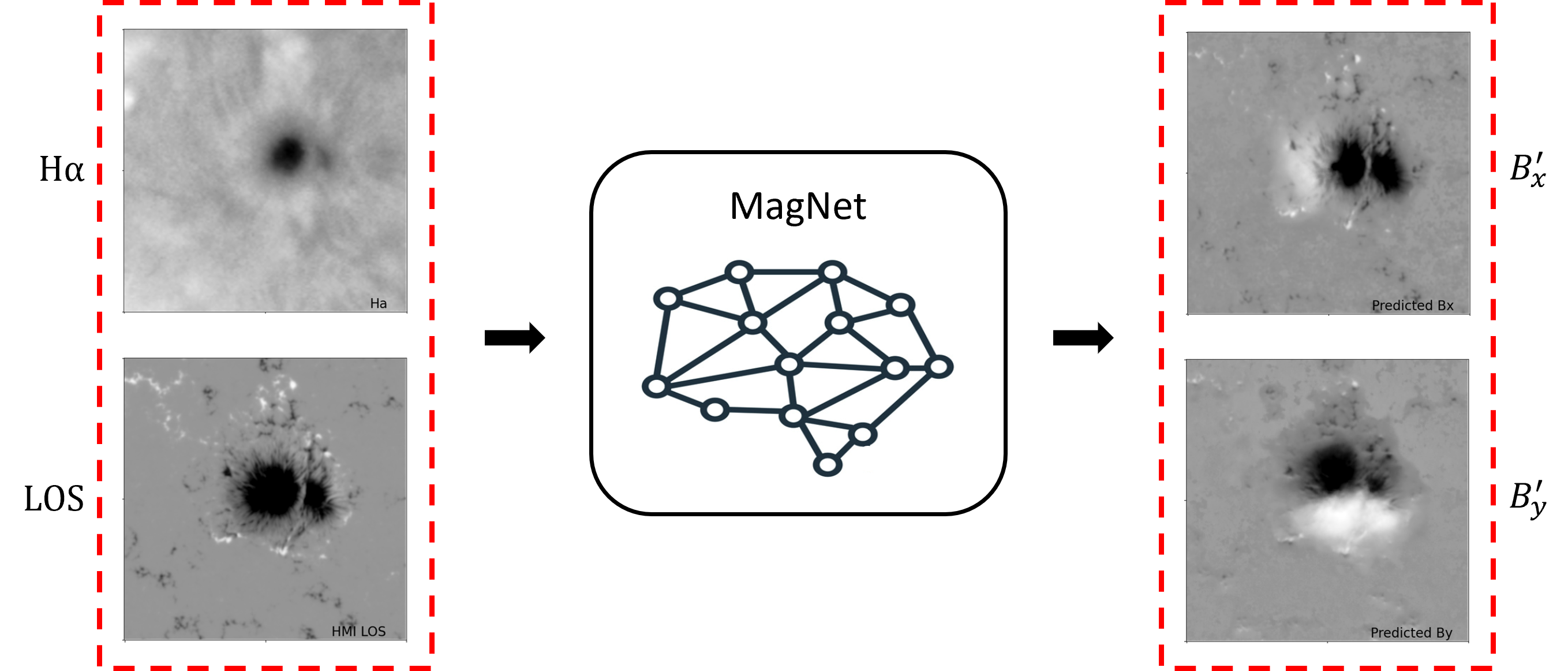}
\caption{MagNet workflow. H$\alpha$  and LOS are MagNet's inputs. B'$_{x}$ and  B'$_{y}$ are fields generated by MagNet. Credit: \cite{Jiang2023}}
\label{workflow}
\vspace{-0.5cm}
\end{figure*}

A schematic diagram of the MagNet workflow is shown in Figure~\ref{workflow}. This machine learning model takes a pair of co-aligned images H$\alpha$ image and the LOS magnetogram as inputs. The outputs are B$_{x}$ and B$_{y}$ of a given solar AR. During the training and testing processes, SDO/HMI vector magnetograms were used as the ``ground truth''. 


SOHO/MDI has spatial resolution is 4'' and the full-disk images are collected on a 1024 × 1024 detector. H$\alpha$, 2032 × 2032 pixels, is collected at 6563~\AA\ wavelength by the 10-inch telescopes at Big Bear Solar Observatory and Kanzelh{\"o}he Solar Observatory \cite{KSO}. H$\alpha$ and SOHO/MDI magnetograms have different pixel resolutions due to the instruments’ differing aperture sizes. The H$\alpha$ images are resized to the MDI's resolution. Use of H$\alpha$ maps allows us to take into account a missing directional component of magnetic fields on the horizontal solar surface direction \cite{Jiang2023}.

In our study, we performed an out-of-sample (OOS) validation of MagNet. OOS validation is an important process in machine learning and statistical models that assesses accuracy of a model using brand new data to the model. Therefore, OOS provides a more realistic and unbiased evaluation in the real world. During the training process, an AI model exposed to a specific dataset and learned within the scope of the dataset. However, in a complex model, the training data might be ``memorized'' together with noise and fluctuations at certain levels. As a consequence, the underlying generalizable patterns
are not fully understood by the model. Such a phenomenon is known as overfitting and it can lead to outstanding performance with training/testing samples, but less accurate on new data. By evaluating the model on a separate OOS data archive which has never been used during the training, we can get a true measure of its ability to generalize. This helps to identify and mitigate overfitting, ensuring that the model's predictions are reliable and robust.




\section{Preparing OOS Data}

The MagNet code takes a pair of co-aligned images, H$\alpha$ + LOS magnetogram (B$_{z}$), as input and predicts the corresponding B$_{x}$ and B$_{y}$ in the same AR as output.  
The key component of OOS validation is to compare the output with new ``ground truth'' data, which is obtained from the Vector Magnetograph (IVM) at Mees Solar Observatory, Haleakala, Maui, Hawaii. 

\subsection{Generating Input Images for MagNet}

Each H$\alpha$-MDI pair contains a full disk H$\alpha$ image and a full disk LOS MDI magnetogram. MDI LOS magnetograms with a cadence of 96 min were accessed through the Joint Science Operations Center (JSOC). For each of the MDI images, we located the H$\alpha$ image obtained nearly simultaneously from Big Bear Solar Observatory (BBSO) and Kanzelhöhe observatory (KSO). 

Since MDI observations are taken in space and therefore are not affected by the turbulence of the Earth’s atmosphere like H$\alpha$ data. We first truncated and aligned H$\alpha$ image to the MDI image, which is used as reference. This alignment is to match the solar disk by their circular shape on both images regardless the magnetic features. After this alignment, all the ARs on disk are automatically aligned with minor shifts, which will be corrected using the cross correlation method, after a certain AR is selected. We used the IDL code align.pro (developed by the University of OSLO / Institute of Theoretical Astrophysics), which aligns two images by utilizing Fourier transform properties to compute their cross-correlation and thereby determine the shift needed to align one image with the reference.

\begin{figure*}[pht]
\centering
\vspace{-4em}
\includegraphics[scale=0.21]{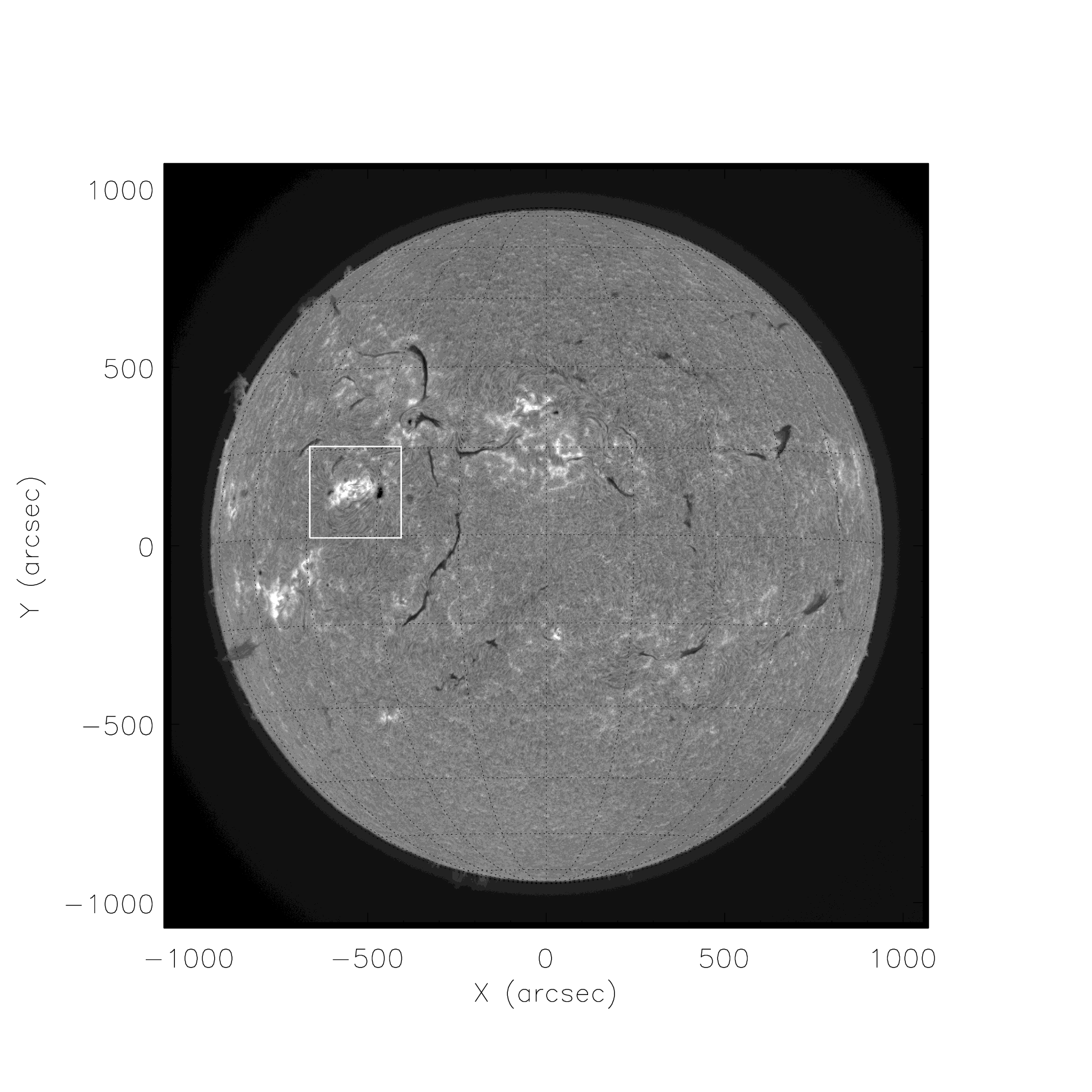}\hspace{0.5em} \vspace{-3.5em}
\includegraphics[scale=0.539]{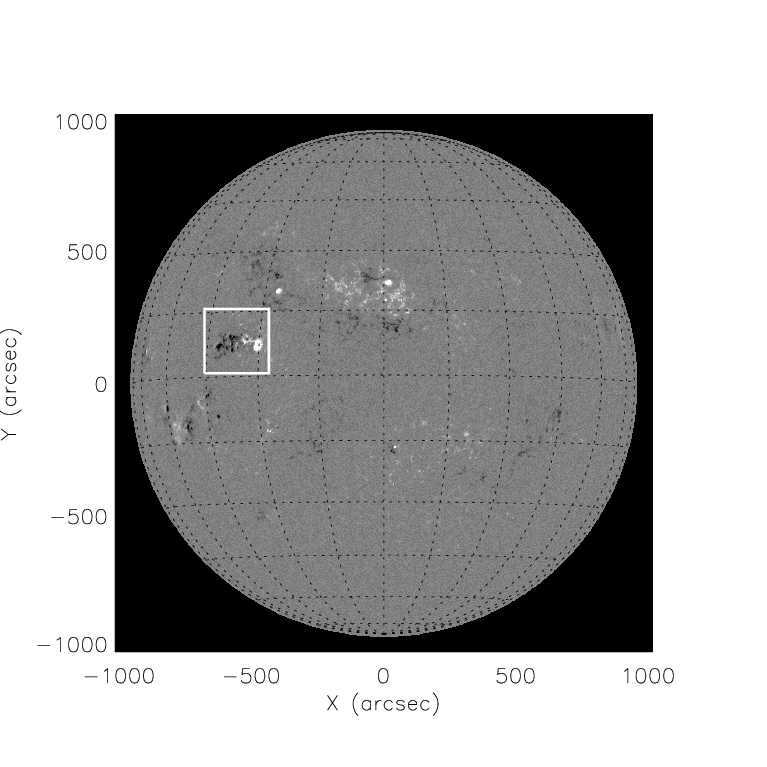}\hspace{-3.5em} \vspace{-0.5em}\\
\includegraphics[scale=0.29]{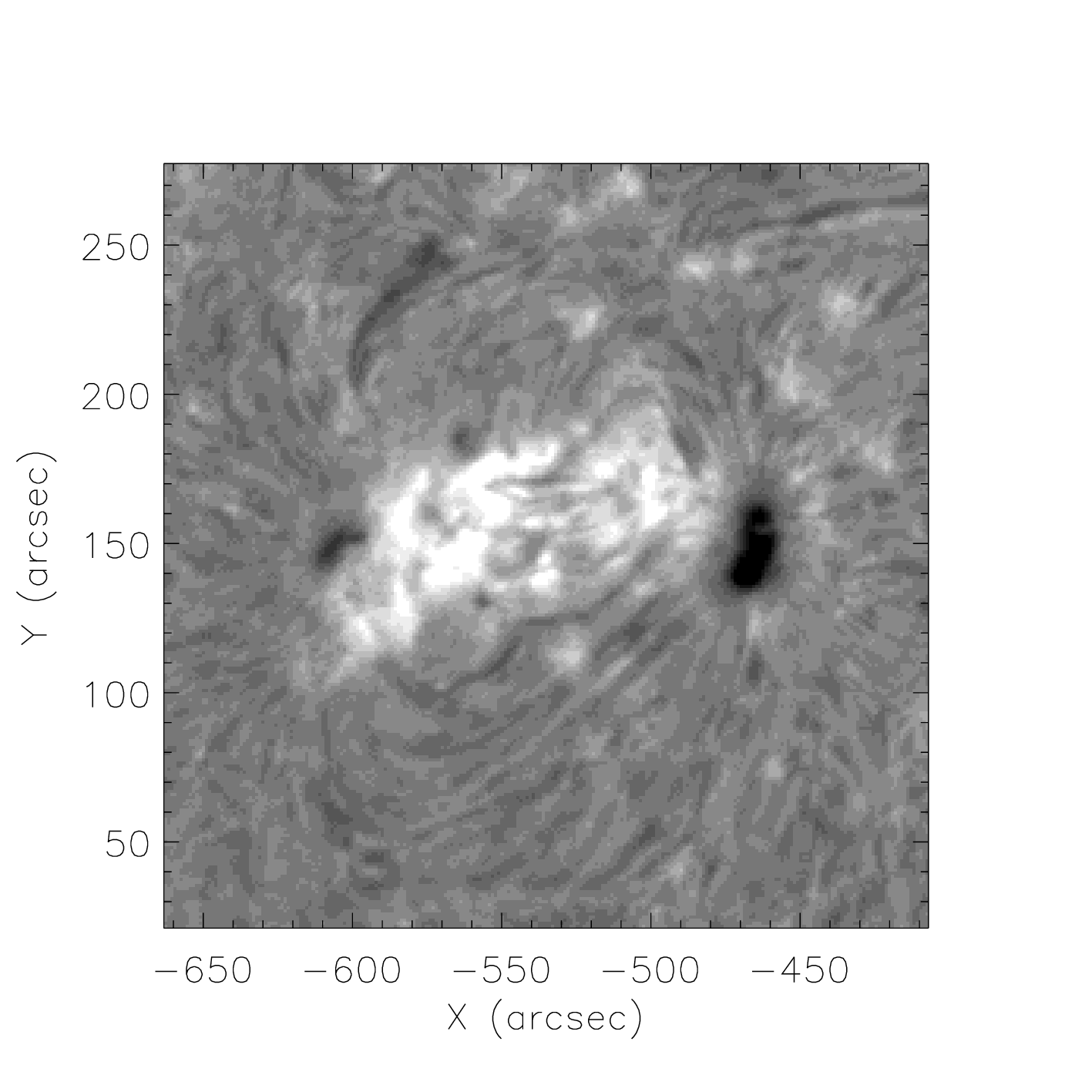}\hspace{0.8em}\vspace{-0.5em}
\includegraphics[scale=0.29]{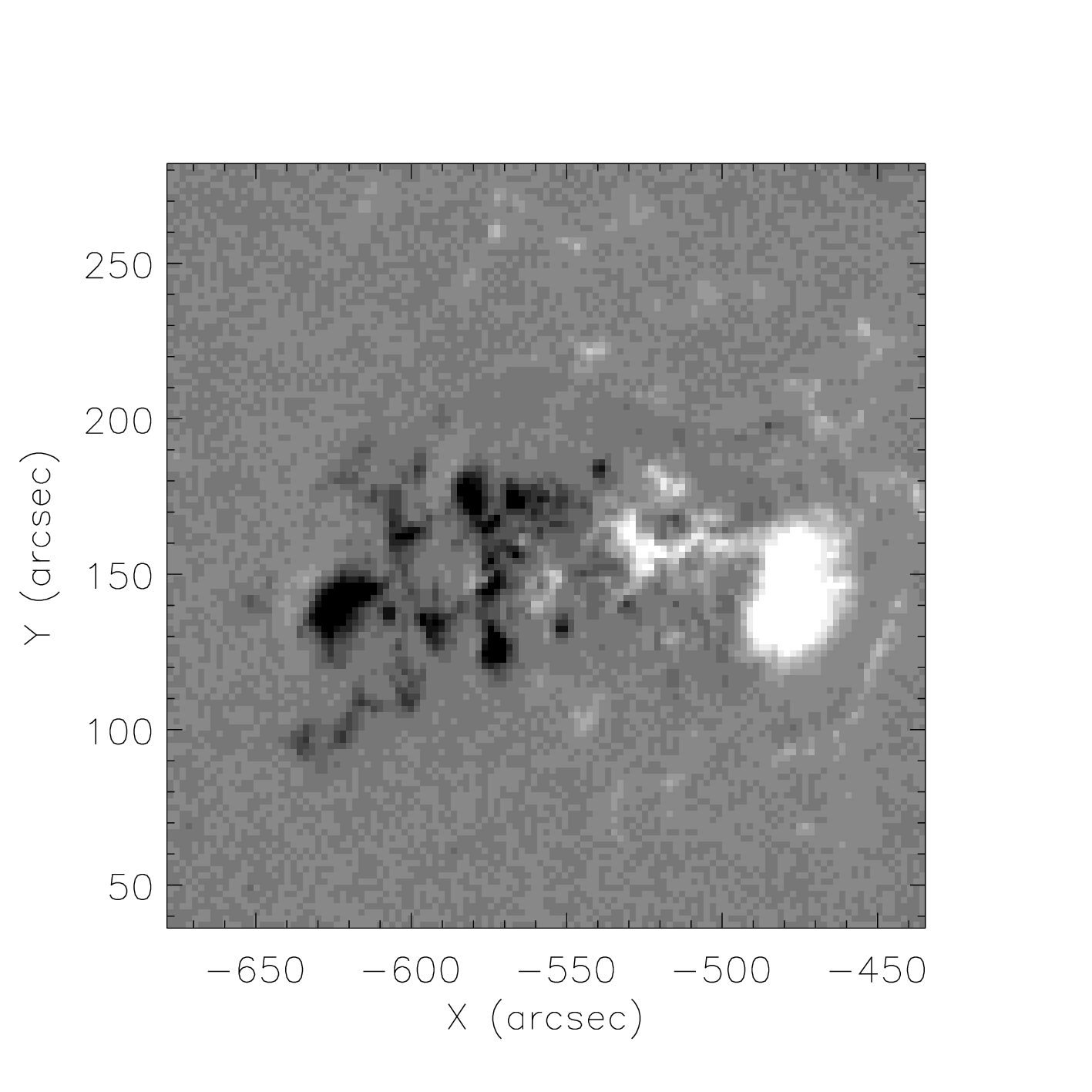}\hspace{-4.5em}
\vspace{-0.5em}
\caption{AR extraction for AR 09643 on 2001-May-21. The left panel shows original full-disk H$\alpha$ image taken at 16:01:04 UT and the selected AR image. Similar maps of the MDI magnetogram taken at 15:59:41 UT are shown in the right panel.}
\label{HaMDI}
\end{figure*}

The location of ARs are identified by the Solar Region Summary (SRS) files, which are daily reports from the Space Weather Prediction Center (SWPC) at National Oceanic and Atmospheric Administration (NOAA). On each day, a NOAA SRS file is published around 00:30 UT. The major contents are the coordinates of ARs or brightenings at that time. Those coordinates can be transferred to other observing times by considering the solar differential rotation (such at the IDL function ``drot\_map.pro'' developed by NASA/GSFC). After the center coordinates of a given AR is determined on the H$\alpha$-MDI pair, we crop the images with a field of view (FOV) of 256 x 256 pixels. As shown in Figure~\ref{HaMDI}, this new image pair with reduced FOV/size is the final inputs for the MagNet model. 





\subsection{MagNet implementation}
\begin{figure*}[pht]
\centering
\vspace{-0.7cm}
\includegraphics[scale=0.85]{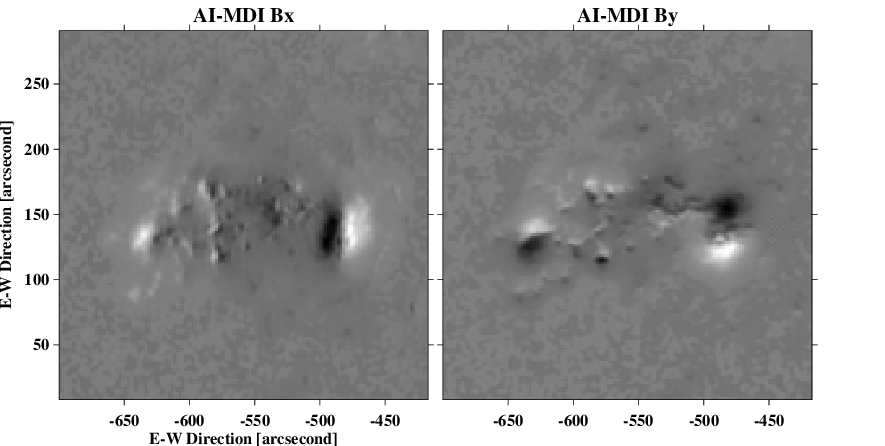}
\caption{AI generated transverse magnetograms, B$_y$(left) and B$_x$(right) of AR 09643 on May 21, 2001.}
\label{ByBxMagNet}
\end{figure*}

MagNet was developed by the Institute for Space Weather Sciences (ISWS) at NJIT. A copy of the trained model has been tested on a workstation in the solar group. Machine learning code was run on the local NJIT server. MagNet reconstructed transverse fields (B$_{x}$ and B$_{y}$ ) are shown in Figure~\ref{ByBxMagNet}.

\section{Validation}
\label{Validation}
In original study \cite{Jiang2023}, the MagNet-generated vector magnetograms was compared with the SDO/HMI vector magnetograms, which is considered as the ``ground truth'',  during the MDI-HMI overlapping period between May 1st, 2010 and April 11, 2011. Here we use the Mees vector magnetograms as the ``ground truth'' to compare with the MagNet-generated vector magnetograms based on MDI LOS magnetograms. We select the AR 09463 observed on May 21, 2001, which was not part of the overlapping period. 

\begin{figure*}[pht]
\centering
\includegraphics[scale=0.88]{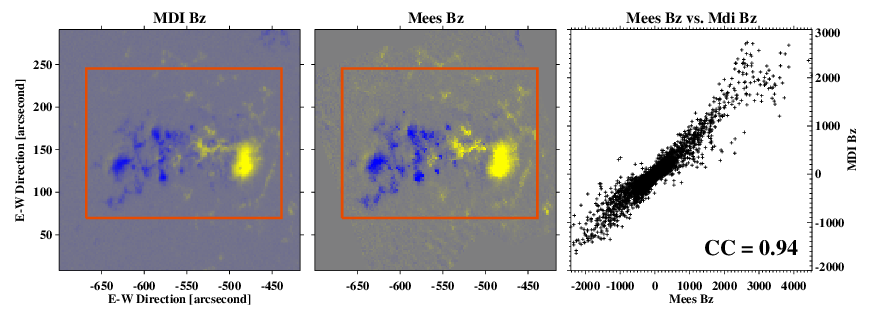}
\includegraphics[scale=0.88]{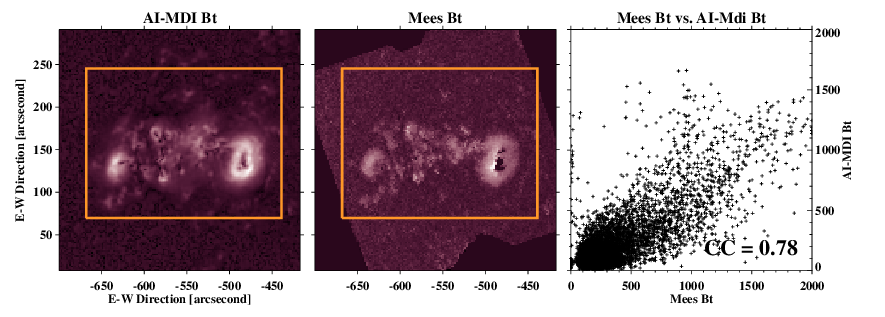}
\caption{Top panels: Comparison between observed MDI B$_z$ with Mees/IVM B$_z$. Bottom panels: Comparison between AI generated vector field B$_t$ with ``ground truth'' Mees/IVM observations.}
\label{validation}
\vspace{-0.2cm}
\end{figure*}

As shown in Figure ~\ref{validation}, the correlation coefficient between MDI B$_z$ and Mees/IVM B$_z$ is 0.94, which is a straightforward result and confirms the consistence between the observed data by MDI and Mees/IVM. The correlation coefficient between the AI generated MDI B$_t$ ($\sqrt{B_x^2 + B_y^2}$) and MEES B$_t$ are shown to be satisfactory: 0.78. Use of transverse fields to validate MagNet's results was intentional, as this method allows us to avoid the effects of 180 degree ambiguity. 180 degree ambiguity refers to uncertainty in the direction of the magnetic field's transverse component. This ambiguity arises from the Zeeman effect, which creates polarized light but does not distinguish between a field pointing in one direction or the opposite along the plane of the Sun's surface. To avoid dealing with this ambiguity, we use transverse field B$_t$, for which question of ambiguity is not relevant. This allows us to ensure that calculated value of the magnetic field is purely a result of MagNet's performance, rather than other unaccounted physical processes. With this OOS validation, the MagMet model can be implemented on all of the data taken in SC23 confidently. 

\section*{Conclusions and Future Plans}
In this project, we were able to successfully generate transverse magnetic fields of AR 09463 from solar cycle 23 using MagNet and perform an OOS validation, which is rarely done. Those findings demonstrated the methodology behind using MagNet and successfully tested its ability to reconstruct missing vector field data for historically significant and magnetically complex regions. When comparing strong magnetic fields, which play a key role in generating eruptions, the model achieved a satisfactory correlation coefficient. 
The MagNet model is currently implemented with ARs in the entire solar cycle 23. Future work will focus on generating additional AR vector maps and comparing the azimuth angle direction between in-sample and out-of-sample validations.


\section*{Acknowledgments}

The authors acknowledge the sponsorship provided by NSF SHINE grant AGS-2228996, NSF CAIG-2425602, NSF’s Research Experience for Undergraduates (REU) Program, 
NASA grant 80NSSC25K7708,
and NJIT’s Undergraduate Translational Research Internship. We thank the science teams of SDO/HMI, SOHO/MDI and Mees/IVM for providing the magnetic data.
The H$\alpha$ data was obtained by Kanzelhöhe Observatory, University of Graz, Austria and Big Bear Solar Observatory, New Jersey Institute of Technology. BBSO operation is supported by US NSF AGS-2309939 grant and New Jersey Institute of Technology. 

\makeatletter
\def\thebibliography#1{\section*{\refname}%
  \normalfont\normalsize
  \list{\@biblabel{\@arabic\c@enumiv}}%
       {\settowidth\labelwidth{\@biblabel{#1}}%
        \leftmargin\labelwidth
        \advance\leftmargin\labelsep
        \usecounter{enumiv}%
        \let\p@enumiv\@empty
        \renewcommand\theenumiv{\@arabic\c@enumiv}}%
  \sloppy
  \clubpenalty4000
  \@clubpenalty \clubpenalty
  \widowpenalty4000%
  \sfcode`\.\@m}
\makeatother

\bibliographystyle{IEEEtran}
\bibliography{./reference, ./newbib}

\end{document}